\pdfoutput=1

\documentclass{article}

\usepackage[a4paper,top=2cm,bottom=2cm,left=3cm,right=3cm,marginparwidth=1.75cm]{geometry}

\usepackage[numbers]{natbib}
\usepackage{bm}
\usepackage{amsmath}
\usepackage{amssymb}
\usepackage{siunitx}
\usepackage{mathtools}
\usepackage{tikz}
\usetikzlibrary{positioning, arrows.meta, calc}
\usetikzlibrary{shapes.geometric}
\usepackage{url}
\usepackage{hyperref}
\usepackage{xspace} 
\usepackage{comment}
\usepackage{float}
\usepackage{url}

\title{A Neural Operator based Hybrid Microscale Model for Multiscale Simulation of Rate-Dependent Materials}

\author{
	Dhananjeyan Jeyaraj\textsuperscript{1}\thanks{Corresponding author: \texttt{jd31@tu-clausthal.de}}, 
	Hamidreza Eivazi\textsuperscript{1}, 
	Jendrik-Alexander Tröger\textsuperscript{2},\\
	Stefan Wittek\textsuperscript{1}, 
	Stefan Hartmann\textsuperscript{2}, 
	Andreas Rausch\textsuperscript{1}
}

\date{
	\textsuperscript{1}Institute for Software and Systems Engineering, Clausthal University of Technology, Germany\\[0.5em]
	\textsuperscript{2}Institute of Applied Mechanics, Clausthal University of Technology, Germany
}

\begin{document}    

\maketitle

\begin{abstract}
	The behavior of materials is influenced by a wide range of phenomena occurring across various time and length scales. To better understand the impact of microstructure on macroscopic response, multiscale modeling strategies are essential. Numerical methods, such as the $\text{FE}^2$ approach, account for micro-macro interactions to predict the global response in a concurrent manner. However, these methods are computationally intensive due to the repeated evaluations of the microscale. This challenge has led to the integration of deep learning techniques into computational homogenization frameworks to accelerate multiscale simulations. In this work, we employ neural operators to predict the microscale physics, resulting in a hybrid model that combines data-driven and physics-based approaches. This allows for physics-guided learning and provides flexibility for different materials and spatial discretizations. We apply this method to time-dependent solid mechanics problems involving viscoelastic material behavior, where the state is represented by internal variables only at the microscale. The constitutive relations of the microscale are incorporated into the model architecture and the internal variables are computed based on established physical principles. The results for homogenized stresses ($<6\%$ error) show that the approach is computationally efficient ($\sim 100 \times$ faster).
\end{abstract}

\vspace{1em}

\section{Introduction}
Modern manufacturing technologies, such as additive manufacturing, enable the design of materials with exceptional mechanical properties \cite{kochmannMultiscaleModelingOptimization2019}. For example, optimizing the topology of a material's microstructure can result in a high strength-to-weight ratio. Another exemplary application is bone implants, where it is essential for the stiffness to match that of the surrounding bone tissue to promote effective bone regeneration \cite{liVitroVivoStudy2016}. Both architectured metamaterials and common engineering materials possess an inherently heterogeneous microstructure, enabling customization of material properties to meet specific engineering demands. As such, developing computational models for these heterogeneous materials is crucial to promote tasks like microstructure optimization, despite the challenges associated with conducting experiments and identifying material parameters \cite{rokosetal2023}.

Modeling the mechanical response of heterogeneous structures presents significant challenges due to the involvement of multiple physical phenomena across different length scales, each contributing to distinct mechanical behaviors. Additionally, the local material behavior is driven by mechanisms that necessitate the identification of appropriate constitutive laws. To address this, multiscale modeling techniques are required, which account for the micro-macro interactions and predict the overall structural response. When both the micro- and macroscales are analyzed using the Finite Element Method (FEM) in a concurrent manner, this approach is referred to as the $\text{FE}^2$ method, see exemplarily \cite{feyelMultiscaleFE2Elastoviscoplastic1999}. In this framework, the macrostructure is discretized spatially using finite elements, and the stress at each integration point is determined through a local finite element simulation using a Representative Volume Element (RVE) representing the heterogeneous microstructure. The resulting local stress is then homogenized and assigned to the macroscale integration point.

However, this multiscale modeling approach is computationally expensive, making its application especially to nonlinear problems nearly prohibitive and rendering the method impractical for nonlinear analyses. This is primarily because nonlinear problems are solved simultaneously at both the micro- and macroscale using iterative methods \cite{tikarrouchineFullyCoupledThermoviscoplastic2019}. Additional factors that can increase computation time include the number of macroscale spatial integration points, the complexity of loading conditions and constitutive behavior, and the mesh resolutions of both RVE and macrostructure. These factors essentially limit the efficient applicability of the $\text{FE}^2$ approach for real-world applications. In contrast, surrogate models based on Machine Learning (ML) and Deep Learning (DL) have demonstrated their effectiveness in reducing computational costs across various applications, including constitutive and multiscale modeling \cite{pengetal2020}.

Several studies have explored the integration of deep learning techniques into computational homogenization frameworks to accelerate multiscale simulations. Deep learning methods, such as Long Short-Term Memory (LSTM) neural networks \cite{danounFELSTMHybridApproach2024}, recurrent neural operators (RNO) \cite{zhangIteratedLearningMultiscale2024}, and deep neural networks (DNN) \cite{eivaziFE2ComputationsDeep2023}, are employed as surrogates for the RVE. The approach involves treating the microscale and macroscale problems separately, eliminating the need for finite element resolution of the microscale problems. Instead, the effective response of the RVE is predicted using a database of offline microscale computations. This approach has resulted in significant reductions in computational time when addressing the microscale problem. When dealing with complex nonlinear inelastic constitutive behavior, internal variables are commonly used in classical constitutive relations. Consequently, various approaches were investigated to handle internal variables in neural network based modeling frameworks \cite{rosenkranzetal2023,rosenkranzetal2024}. Unlike the aforementioned studies that focus on substitutive methods, a complementary approach was introduced in \cite{eivazi2025equino} for quasi-static problems. Therein, a hybrid method was proposed incorporating kinematic and constitutive relations while inherently satisfying balance relations at the microscale by construction.

In this work, we present a hybrid modeling approach for predicting microscale physics within multiscale simulations using operator learning techniques. Our method incorporates known physical principles directly into the framework, including kinematic and constitutive relations as well as the ordinary differential equations governing the evolution of internal variables. While similar strategies have been applied in other contexts, their use in rate-dependent simulations remains largely unexplored. We build upon deep operator networks (DeepONets) as introduced by \cite{luLearningNonlinearOperators2021}, extending them to account for history dependence through latent variables. Homogenized quantities are computed in a manner analogous to traditional numerical homogenization methods. The proposed model is applied to rate-dependent solid mechanics problems exhibiting viscoelastic behavior. Results demonstrate that our method achieves high accuracy and efficiency, even when trained on relatively small datasets. Furthermore, we show that a high-performance implementation leveraging modern computing libraries and just-in-time (JIT) compilation can achieve speedups of several orders of magnitude compared to conventional $\text{FE}^2$ approaches.

The structure of this manuscript is as follows. The next section presents the neural operator-based hybrid model, including a discussion on operator learning and the adopted methodology. This is followed by a numerical example that illustrates the effectiveness of the proposed approach, with details on the problem setup, RVE simulation, and results. The final section concludes the manuscript with a summary and outlines directions for future work.

\section{Neural Operator based hybrid model}
We begin by introducing the neural operator-based hybrid modeling framework, emphasizing the concepts behind operator learning and the methodology used in this study. 

\subsection{Operator Learning}
Approximation of nonlinear functionals defined on some function space is often a problem in engineering applications. This may involve approximating nonlinear operators that map from one infinite-dimensional function space to another. The authors in \cite{chenUniversalApproximationNonlinear1995} developed a theorem that any nonlinear continuous operator can be approximated with a neural network with a single hidden layer. However, this theorem only guarantees a small approximation error without considering the optimization error and generalization error. To overcome this problem, DeepONets \cite{luDeepONetLearningNonlinear2019} were developed to learn operators accurately and efficiently from small datasets. A trained DeepONet can be applied to different input functions to produce results faster than numerical solvers and can be applied to simulation data, experimental data or both \cite{higginsGeneralizingUniversalFunction2021}. Another widely used method in operator regression is the Fourier Neural Operator (FNO), which, in its continuous formulation, can be interpreted as a specific instance of a DeepONet, where both the branch and trunk networks are structured using a trigonometric basis \cite{kovachkiUniversalApproximationError2021}. A detailed comparison of these techniques, along with their respective limitations, is provided in \cite{luComprehensiveFairComparison2022}, which also explores some DeepONet extensions where the basis functions are derived via proper orthogonal decomposition (POD). Additional operator learning strategies involve integrating neural networks with model reduction techniques \cite{bhattacharyaModelReductionNeural2021,eivazi2024nonlinear}. In such approaches, model reduction is employed to identify a low-dimensional approximation space, and learning occurs within this latent space. Another viewpoint involves learning an operator based on its response to a representative set of input functions \cite{patelPhysicsinformedOperatorRegression2021}, framing the problem as a reduced-space constrained optimization task.

\subsection{Methodology}
\begin{figure*}
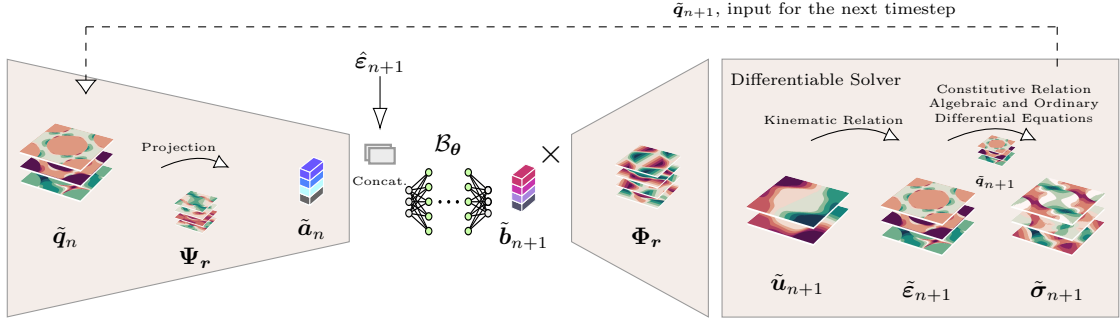

	\centering
    \include{algo2}
	\caption{An overview of the proposed methodology}
	\label{FIG:1}
\end{figure*}

An overview of the methodology followed in this work is shown in Figure~\ref{FIG:1}. Finite elements are utilized to spatially discretize the heterogeneous microscale domain, which is represented by an RVE \cite{staubMultiScaleSimulationViscoelastic2012}. The deformation behavior at the boundaries of the RVE is driven by applying periodic displacement boundary conditions using the macroscale strain $\hat{\bm{\varepsilon}}^j$ of the corresponding macroscale integration point $j$. Note that we consider only two-dimensional structures in this study, however, the general methodology is applicable in three dimensions as well.  Let $\Omega$ denote the microscale domain of interest, assumed to be a symmetric and zero-centered unit cell, such that
\begin{equation}
\Omega = \left\{ \bm{x} = \{x_d\} \in \mathbb{R}^d \,\middle|\, -\frac{L}{2} \leq x_d \leq \frac{L}{2} \right\},
\end{equation}

\noindent
for any point $\bm{x}$, where $L$ denotes the edge length of $\Omega$ and $d$ represents the spatial dimension of the problem \cite{henkeswesselsmahnken2022}. The balance of linear momentum (without body forces and dynamic terms) represents the governing balance relation in this work,
\begin{equation}
\bm{\sigma}(\bm{x},t)\nabla = \bm{0}, \quad \bm{x}\in\Omega,
\end{equation}

\noindent
following the notation in \cite{haupt2002}. The stress tensor $\bm{\sigma}$ is obtained from an inelastic constitutive relation,
\begin{equation}
\label{eq:constitutiveRelation}
\bm{\sigma}(\bm{x},t) = \bm{h}(\bm{\varepsilon},\bm{q}),
\end{equation}

\noindent
where $\bm{\varepsilon}$ denotes the linearized strain tensor and $\bm{q}\in\mathbb{R}^{n_q}$ are the internal variables that develop according to evolution equations
\begin{equation}
\label{eq:evolutionEq}
\dot{\bm{q}}(\bm{x},t) = \bm{r}_q(\bm{\varepsilon},\bm{q}).
\end{equation}

\noindent
In the following, we will formulate displacements and internal variables as column vectors since we consider these quantities as degrees of freedom at specific locations in $\Omega$. Similarly, the stresses and strains are consistently formulated as column vectors employing the symmetry of the tensors as it is common in the context of numerical methodologies. 

The resulting initial boundary-value problem at the microscale is solved using a hybrid modeling approach. The objective is to learn the solution operator that governs the microscale mechanical response under varying periodic boundary conditions, which are induced by the prescribed macroscale strain $\hat{\bm{\varepsilon}}(t)$. The index $j$ is omitted for brevity. This is done while accounting for the evolving state of the microstructure, represented by the internal variables $\bm{q}(t_n) = \bm{q}_n$. The central idea is to employ a neural network to approximate the displacement field $\bm{u}_{n+1}(\bm{x}, \hat{\bm{\varepsilon}}_{n+1}, \bm{q}_{n+1})$ at any spatial location $\bm{x} \in \Omega$, as the system advances from time step $t_n$ to $t_{n+1}$. The model conditions this prediction on the internal variables $\bm{q}_n$ at time $t_n$, thereby capturing the microscale behavior.

A DeepONet approximates this field by employing two neural networks: a branch network and a trunk network \cite{luLearningNonlinearOperators2021}. The branch network receives the macroscale strain $\hat{\bm{\varepsilon}}$ and internal variables $\bm{q}_{n}$ as inputs and outputs the coefficients corresponding to a set of basis functions. These basis functions are simultaneously learned for the output function $\bm{u}$ from the training data. However, we use POD-DeepONet \cite{luComprehensiveFairComparison2022} in this work as it outperforms DeepONet in most of the benchmark cases. In this architecture, the trunk network is replaced by a set of precomputed POD bases. Moreover, the internal variables $\bm{q}_{n}$ are high dimensional. The idea here is to use a latent representation of the internal variables by projecting them onto the POD modes of the internal variables. The snapshot matrices $\bm{U}$ and $\bm{Q}$ are constructed from a limited set of output function samples evaluated at spatial collocation points $\bm{\xi} \in \mathbb{R}^m$ within the domain $\Omega$, where $m = n_\text{comp} \cdot n_\text{coll}$ ($n_\text{comp}$ denotes the number of components in the field and $n_\text{coll}$ the number of collocation points), and at $T$ discrete time points in the temporal domain for each of the $N$ trajectories, as follows:
%
\begin{equation}
    \bm{U} = [\bm{u}_1^1, \ldots,\bm{u}_T^1, \ldots, \bm{u}_T^N],
\end{equation}    
and
\begin{equation}
    \bm{Q} = [\bm{q}_1^1, \ldots,\bm{q}_T^1, \ldots, \bm{q}_T^N],
\end{equation}    
\noindent
respectively, resulting in the POD basis functions or the so-called modes. We refer to $\bm{u}_n^i(\bm{\xi}, \hat{\bm{\varepsilon}}_n^i, \bm{q}_n^i)$ as $\bm{u}_n^i$ and $\bm{q}_n^i(\bm{\xi}, \bm{\varepsilon}_n^i, \bm{q}_{n-1}^i)$ as $\bm{q}_n^i$ for brevity, where $i = 1,\ldots,N$ indicates the trajectory and $n=1,\ldots,T$ is the time step in each trajectory. This results in snapshot matrices of shape $m \times s$, where $m$ is the number of spatial degrees of freedom and $s = T \cdot N$ denotes the total number of snapshots across all trajectories. We perform reduced singular value decomposition (SVD) on the snapshot matrix as follows
\begin{equation}
\bm{{Q}} \approx \bm{\Psi}_r \bm{\Sigma}_r \bm{V}_r^\mathsf{T},
\label{eq:svd_reduced}
\end{equation}
where $\bm{\Psi}_r \in \mathbb{R}^{m \times r}$ contains the leading $r$ left singular vectors (POD modes), $\bm{\Sigma}_r \in \mathbb{R}^{r \times r}$ is the diagonal matrix of the $r$ largest singular values, and $\bm{V}_r^\mathsf{T} \in \mathbb{R}^{r \times s}$ is the truncated transpose of the right singular matrix. Similarly, performing a reduced SVD on the snapshot matrix $\bm{{U}}$ yields the POD modes $\bm{\Phi}_r$. For given internal variables $\bm{q}_n(\bm{\xi}, \bm{\varepsilon}_n, \bm{q}_{n-1})$ at time $t_n$, the projection coefficients $a_{i,n}$ are computed by projecting the internal variables onto the reduced POD basis
\begin{equation}
a_{i,n} = \langle \bm{q}_n(\bm{\xi}, \bm{\varepsilon}_n, \bm{q}_{n-1}), \bm{\psi}_i \rangle, \quad i = 1, \dots, r,
\label{eq:pod_coeff}
\end{equation}
where $\bm{\psi}_i$ denotes the $i$-th column of $\bm{\Psi}_r$. Collecting these coefficients into a vector, we define
\begin{equation}
\bm{a}_n := \{a_{i,n}\}_{i=1}^r.
\label{eq:coeff_vector}
\end{equation}
\noindent
The displacement vector $\bm{u}_{n+1}(\bm{x}, \hat{\bm{\varepsilon}}_{n+1}, \bm{q}_{n+1})$ at any point $\bm{x} \in \Omega$ is approximated by the operator network $\mathcal{N}_{\bm{\theta}}$ and written as
\begin{equation}
\begin{split}
\bm{u}_{n+1}(\bm{x}, \hat{\bm{\varepsilon}}_{n+1}, \bm{q}_{n+1}) \approx \mathcal{N}_{\bm{\theta}}(\bm{x}, \hat{\bm{\varepsilon}}_{n+1}, \bm{a}_n) \\
= \sum_{k=1}^{r} \mathcal{B}_{\bm{\theta}, k}(\hat{\bm{\varepsilon}}_{n+1}, \bm{a}_n; \bm{\theta})\, \bm{\phi}_k(\bm{x}) + \bm{\phi}_0(\bm{x}),
\end{split}
\label{eq:neural_operator}
\end{equation}
where the predicted coefficient vector is obtained from the branch network $\mathcal{B}_{\bm{\theta}}$,
\begin{equation}
\tilde{\bm{b}} = \mathcal{B}_{\bm{\theta}}(\hat{\bm{\varepsilon}}_{n+1}, \bm{a}_n; \bm{\theta}),
\label{eq:branch_net}
\end{equation}
with $\tilde{\bm{b}} = [\tilde{b}_1, \tilde{b}_2, \ldots, \tilde{b}_r]^\mathsf{T}$ and $\bm{\theta}$ is the collection of all trainable weights and bias parameters in the model. The functions $\bm{\phi}_k(\bm{x})$ are the POD modes of the displacement field, and $\bm{\phi}_0(\bm{x})$ denotes the mean field. The branch network is trained by minimizing the following loss function:
\begin{equation}
\mathcal{L}_{\text{branch}}(\bm{\theta}) 
= \frac{1}{s} \sum_{j=1}^{s}
{(\tilde{\bm{b}}^{(j)} - \bm{b}^{(j)})^2},
\label{eq:branch_loss}
\end{equation}
where $\bm{b}^{(j)}$ and $\tilde{\bm{b}}^{(j)}$ denote the true and predicted coefficient vectors, respectively, for the input samples of macroscopic strain $\hat{\boldsymbol{\varepsilon}}$ drawn from the set $\boldsymbol{\mathcal{E}}$. The reference coefficients $\bm{b}^{(j)}$ are obtained by projecting the displacement field from the training data onto the reduced POD basis $\bm{\Phi}_r$. Therefore, this approach combines a surrogate strategy—where a data-driven model replaces the iterative solution procedure of physics-based models—with an assistant strategy, in which outputs from a physics-based model serve as additional inputs to the data-driven model \cite{krannichfeldtCombiningPhysicsbasedDatadriven2025}. The microscale strain tensor is obtained from the displacement vector as follows
%
\begin{equation}
\begin{split}
\bm{\varepsilon}(\bm{u}) = \dfrac{1}{2} (\bm{u} \otimes \nabla + \nabla \otimes \bm{u}),
\end{split}
\end{equation}

\noindent
using the linearized kinematic relation. In the proposed architecture, the strain tensor is computed by projecting the kinematic relation onto the truncated POD modes, in a manner analogous to Galerkin methods
\begin{align}
\begin{split}
& \boldsymbol{\varepsilon}_{n+1}(\boldsymbol{x}, \hat{\boldsymbol{\varepsilon}}_{n+1}, \bm{a}_n) \approx \\ & \frac{1}{2} \Bigg[ \sum_{k=1}^{r} \mathcal{B}_{\bm{\theta}, k}(\hat{\boldsymbol{\varepsilon}}_{n+1}, \bm{a}_n; \bm{\theta})
\left( \nabla \bm{\phi}_k(\bm{x}) + \nabla \bm{\phi}_k^\mathsf{T}(\bm{x}) \right) \\ 
& \hspace{3cm} + (\nabla \bm{\phi}_0(\bm{x}) + \nabla \bm{\phi}_0^\mathsf{T}(\bm{x})) \Bigg].
\end{split}
\label{eq:time_dependent_strain}
\end{align}
The evolution of the internal variables according to the evolution equation \eqref{eq:evolutionEq} and the constitutive relation \eqref{eq:constitutiveRelation} are employed to describe the stress-strain dependence. In this work, we rely on established and thus thermodynamically consistent constitutive relations in contrast to data-driven constitutive modeling approaches. Nevertheless, data-driven constitutive relations could be incorporated into our framework as well.

\section{Numerical Example}
To evaluate the performance of the proposed hybrid model, we present a numerical example that illustrates its application and effectiveness.

\subsection{Problem Setup}
\begin{figure}[ht!]
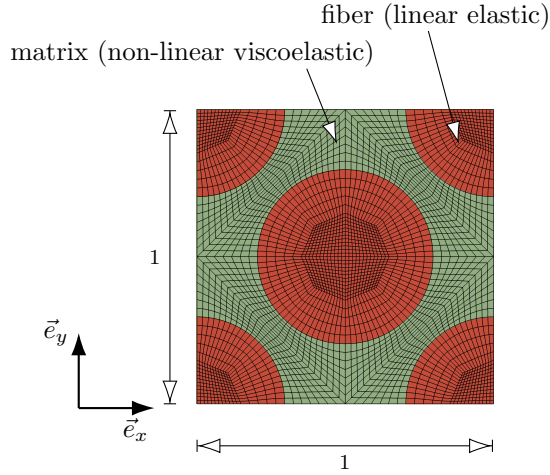

	\centering
    \include{rve}
	\caption{RVE at microscale}
	\label{FIG:2}
\end{figure}

In this study, we consider a two-dimensional RVE inspired by fiber-reinforced polymer materials, see Figure~\ref{FIG:2}, assuming plane strain state. 
The fibers are modeled as linearly elastic 
\begin{equation}
\bm{\sigma} = K_f \mathrm{tr}(\bm{\varepsilon})\bm{I} + 2 G_f \bm{\varepsilon}^\mathrm{D},
\end{equation}
with typical material parameters of glass fibers, $K_f = \SI{4.35e04}{\MPa}$ and $G_f = \SI{2.99e04}{\MPa}$. Note that $\bm{\varepsilon}^\mathrm{D}$ indicates the deviatoric strain tensor. The polymer matrix is assumed to behave nonlinear viscoelastic, which is based on an additive decomposition of the stresses and strains,
\begin{equation}
\bm{\sigma} = \bm{\sigma}_\mathrm{eq} + \bm{\sigma}_\mathrm{ov} \quad \text{and} \quad \bm{\varepsilon} = \bm{\varepsilon}_\mathrm{e} + \bm{\varepsilon}_\mathrm{v}.
\end{equation}
The specific relations for the equilibrium stresses and overstresses read
\begin{align}
\bm{\sigma}_\mathrm{eq} &= K_m \mathrm{tr}(\bm{\varepsilon})\bm{I} + 2 G_m \bm{\varepsilon}^\mathrm{D}, \\
\bm{\sigma}_\mathrm{ov} &= 2\hat{G}(\bm{\varepsilon}-\bm{\varepsilon}_\mathrm{v})^\mathrm{D}.
\end{align}
Note that the overstresses are deviatoric, $\bm{\sigma}_\mathrm{ov} = \bm{\sigma}_\mathrm{ov}^\mathrm{D}$. The viscous strains evolve according to an ordinary differential equation
\begin{equation}
\label{eq:viscousStrains}
\dot{\bm{\varepsilon}}_\mathrm{v} = \frac{1}{\eta(\bm{\sigma})}\bm{\sigma}_\mathrm{ov} = \frac{2\hat{G}}{\eta(\bm{\sigma})}(\bm{\varepsilon}-\bm{\varepsilon}_\mathrm{v})^\mathrm{D},
\end{equation}
where the nonlinearity is introduced by the stress-dependent viscosity function
\begin{equation}
\eta(\bm{\sigma}) = \eta_0 \exp(-s_0 \vert\vert \bm{\sigma}_\mathrm{ov} \vert\vert).
\end{equation}
Here, Eq.~\eqref{eq:viscousStrains} is the specific evolution equation in the sense of Eq.~\eqref{eq:evolutionEq}. Accordingly, the viscous strains $\bm{\varepsilon}_\mathrm{v}$ are the internal variables $\bm{q}$ at each spatial integration point of the microstructure. In two-dimensional problems, $\bm{\varepsilon}_\mathrm{v}\in\mathbb{R}^4$ holds. The nonlinear model comprises five material parameters, $K_m = \SI{2778}{\MPa}$, $G_m = \SI{185.2}{\MPa}$, $\hat{G} = \SI{740.7}{\MPa}$, $\eta_0 = \SI{1.3e05}{\MPa\s}$, and $s_0 = \SI{1.1}{\per\MPa}$. The nonlinear viscoelasticity represents a differential-algebraic system of equations at each spatial integration point. It is noteworthy that an efficient reduction to a scalar nonlinear equation is possible, in contrast to solving a system of nonlinear equations (after time-discretizing Eq.~\eqref{eq:viscousStrains} using Backward Euler method), see \cite{hartmann2005} for details. The scalar nonlinear equation is solved using the Pegasus method in this work.

\subsection{RVE simulation}
\begin{figure*}[ht!]
	\centering
        \includegraphics[width=\textwidth]{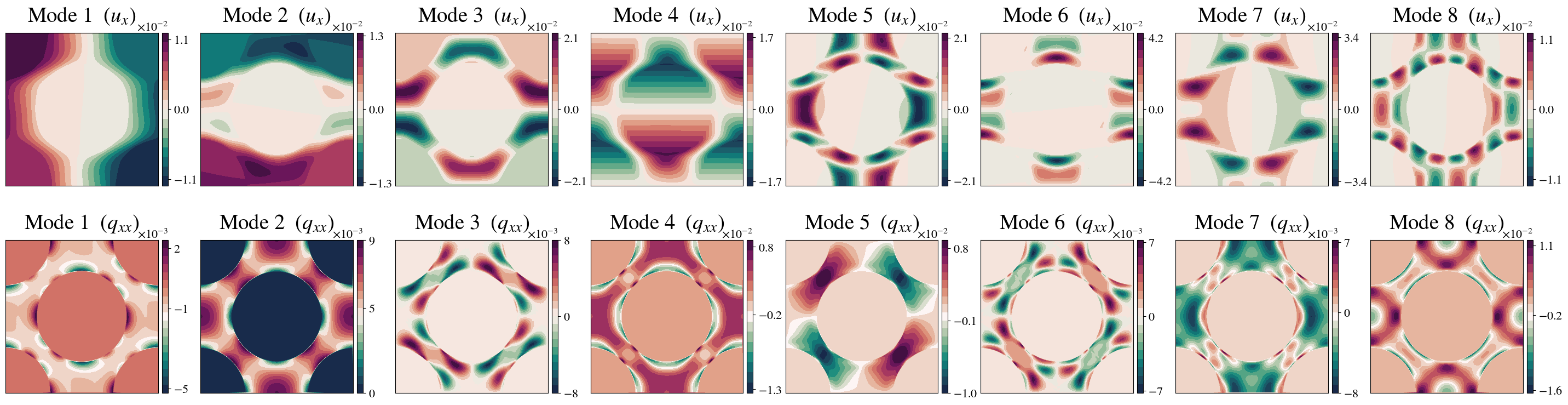}
	\caption{First 8 dominant modes of $u_x$ and $q_{xx}$}
	\label{FIG:3}
\end{figure*}
As the hybrid model includes a data-driven component approximated by a POD-DeepONet, training data is necessary. This data is generated by conducting RVE simulations using finite elements. The inputs to these simulations are macroscale strains sampled within a magnitude of $\pm 0.04$ in tension and compression for each strain component, using Latin Hypercube Sampling (LHS) \cite{mckayComparisonThreeMethods1979}. We prescribe a nonlinear evolution of the macroscale strain according to
\begin{equation}
\hat{\boldsymbol{\varepsilon}}(t) = A \sin(\omega t), \quad \omega = 2 \pi \times \SI{0.05}{\per\s},
\label{eq:strain_time}
\end{equation}
where the amplitude $A$ is sampled via LHS from a uniform distribution $\mathcal{U}(-0.04, 0.04)$, and the discrete time points are sampled over the interval
\[
t \in [0, 10]\si{\s}, \quad \text{with increments of } \Delta t = \SI{0.1}{\s}.
\]
Training data is generated from 1000 such random inputs resulting in 1000 trajectories each containing 100 discrete time steps. It should be noted that, of course, the macroscale strain will be obtained from macroscale finite element simulations as soon as the hybrid microscale model is integrated into a multiscale framework leveraging finite elements and operator learning. Since this is left for future work, we prescribe the macroscale strain in the following. The simulation outputs used for the POD computations include the displacement field and internal variables. Since the number of POD modes is a hyper-parameter, we analyze the cumulative energy to determine how many modes are needed for accurate reconstruction. The results show that the first 16 POD modes suffice to reconstruct both the displacement field and internal variables accurately. Figure~\ref{FIG:3} illustrates the first 8 dominant modes of the displacement in the $x$-direction and internal variable, i.e.\ the viscous strain, in the $x$-direction.

To train the POD-DeepONet, the dataset was divided into training (90\%) and test (10\%) subsets. 20\% of the training data was used for validation. The branch network in the proposed architecture consists of two hidden layers, each with 128 neurons, and utilizes the swish activation function. Training was performed with a batch size of 64. The mean-squared error, as in Eq.~\eqref{eq:branch_loss}, was employed as the loss function and optimized using the Adam optimizer. The learning rate followed a scheduled exponential decay, starting at 0.001 with decay steps of 1000 and a decay rate of 0.5. All training computations were carried out on an NVIDIA GeForce RTX 3090 GPU. A trained neural network is used to predict the displacement field and then evaluated using the following metric:

\begin{equation}
\text{Mean relative } \mathcal{\ell}_2 \text{ norm of error} = \frac{1}{N_\text{test}} \sum_{i=1}^{N_\text{test}} \frac{\| \bm{u}^{(i)} - \tilde{\bm{u}}^{(i)} \|_2}{\| \bm{u}^{(i)} \|_2},
\label{eq:mean_rel_l2_error}
\end{equation}
where $N_\text{test}$ is the number of test samples, $\bm{u}^{(i)}$ denotes the true displacement field of the $i$-th test sample obtained from the reference solution, and $\tilde{\bm{u}}^{(i)}$ represents the predicted displacement field from the operator network for the same sample. The two-dimensional displacement field ($n_\text{comp} = 2$) is discretized with $n_\text{coll} = 10,562$ collocation points, which correspond to the finite element nodes of the spatial discretization visualized in Figure~\ref{FIG:2} using 8-noded quadrilateral elements. The internal variable field is spatially discretized with $n_\text{coll} = 31,104$ collocation points and has 4 components. Note that the collocation points are chosen as the standard integration points of the 8-noded quadrilateral elements when employing Gauss-Legendre quadrature in two dimensions. Both stress and strain field are discretized with $n_\text{coll} = 31,104$ collocation points as well and comprise $n_\text{comp} = 4$ and $n_\text{comp} = 3$ components, respectively.

\subsection{Results}
\label{sec:results}
\begin{figure}[htbp!]
	\centering
	\includegraphics[width=.8\columnwidth]{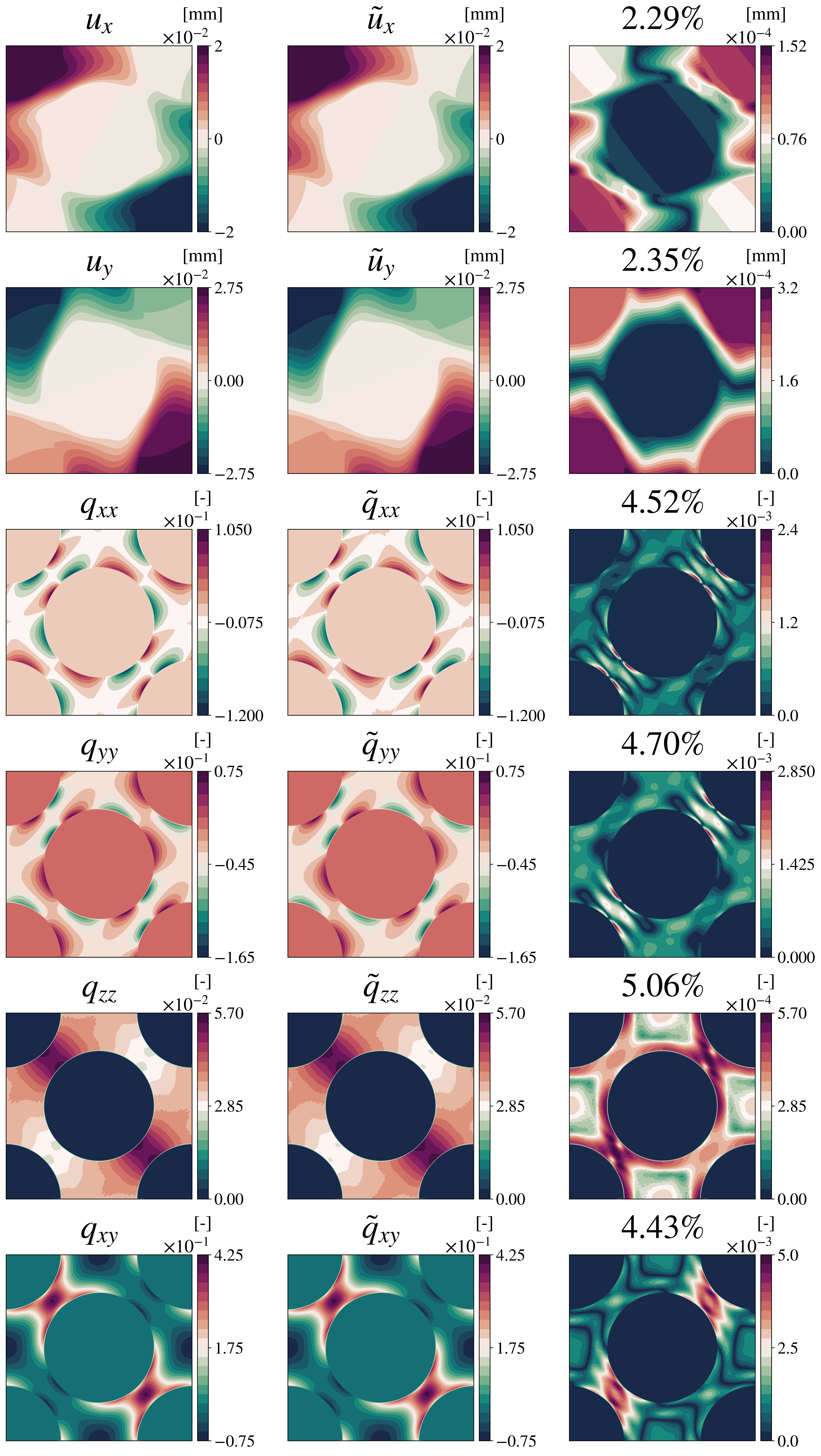}
	\caption{Microscale displacements and internal variables for input strain $\hat{\bm{\varepsilon}} = [-0.02151,\ -0.03329,\ -0.00358]^\mathsf{T}$ at $t = \SI{5}{\s}$. Columns show reference, prediction, and absolute error, respectively. Titles in the third column indicate the relative $\ell_2$ norm of error for the test dataset.}
	\label{FIG:4}
\end{figure}

A total of 100 trajectories, completely unseen during training, each consisting of ten time steps, were analyzed. The error analysis was conducted on these individual samples for each predicted quantity, comparing them to the reference values using the metric described in Eq.~\eqref{eq:mean_rel_l2_error}. A more rigorous approach to analyzing the results involves modeling the temporal evolution of internal variables such that the state at each time step is conditioned on the preceding states, allowing for the dynamics to propagate consistently across the entire trajectory. Performing such analysis on the test dataset, the displacement field exhibited errors of approximately 2.29\% and 2.35\% in the $x$- and $y$-directions, respectively. These results correspond to the data-driven component of the hybrid model. Using the predicted displacement field, the microscale strain field was computed based on known physical principles, specifically the kinematic relation~\eqref{eq:time_dependent_strain}. The error analysis revealed deviations of about 3.02\%, 2.95\%, and 3.36\% in the $\varepsilon_{xx}$, $\varepsilon_{yy}$, and $\varepsilon_{xy}$ components of the strain field, respectively.

\begin{figure}[htbp!]
	\centering
	\includegraphics[width=.8\columnwidth]{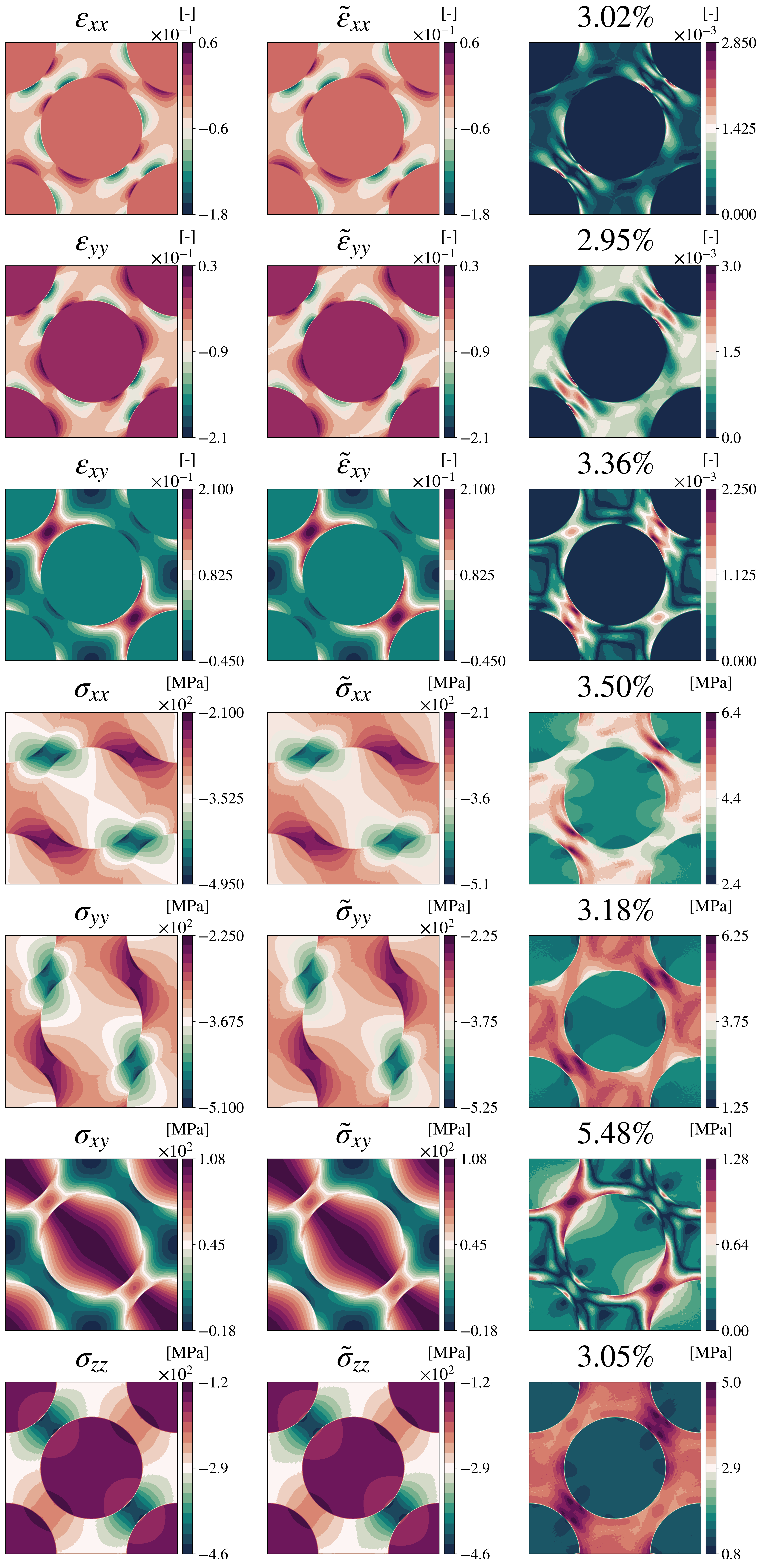}
	\caption{Microscale strains and stresses for input strain $\hat{\bm{\varepsilon}} = [-0.02151,\ -0.03329,\ -0.00358]^\mathsf{T}$ at $t = \SI{5}{\s}$. Columns show reference, prediction, and absolute error, respectively. Titles in the third column indicate the relative $\ell_2$ norm of error for the test dataset.}
	\label{FIG:5}
\end{figure}
Subsequently, the internal variables were calculated by solving the governing ordinary differential equations describing their temporal evolution. The deviations in the internal variable components compared to the reference values were 4.52\%, 4.70\%, 4.43\%, and 5.06\% for the $q_{xx}$, $q_{yy}$, $q_{xy}$ and $q_{zz}$ components, respectively. Finally, the constitutive relation was employed to compute the microscale stresses. Deviations of about 3.50\%, 3.18\%, 5.48\%, and 3.05\% were observed for the $\sigma_{xx}$, $\sigma_{yy}$, $\sigma_{xy}$, and $\sigma_{zz}$ components in the test dataset. These results support the viability of the proposed multiscale framework. In particular, the accuracy of the strain field is critical for homogenization and information transfer to the macroscale, since internal variables are retained and evolved at the microscale. Furthermore, conventional numerical homogenization techniques, i.e.\ volume integrals, were used to compute the homogenized strain $\hat{\boldsymbol{\varepsilon}}(t)$ and stress $\hat{\boldsymbol{\sigma}}(t)$. Figure~\ref{FIG:4} and Figure~\ref{FIG:5} shows the predicted simulation results for microscale RVE for one example trajectory with median error and macroscale strain $\hat{\bm{\varepsilon}} = [-0.02151,\ -0.03329,\ -0.00358]^\mathsf{T}$ at $t=\SI{5}{\s}$.
\begin{figure}[htbp!]
	\centering
	\includegraphics[width=.8\columnwidth]{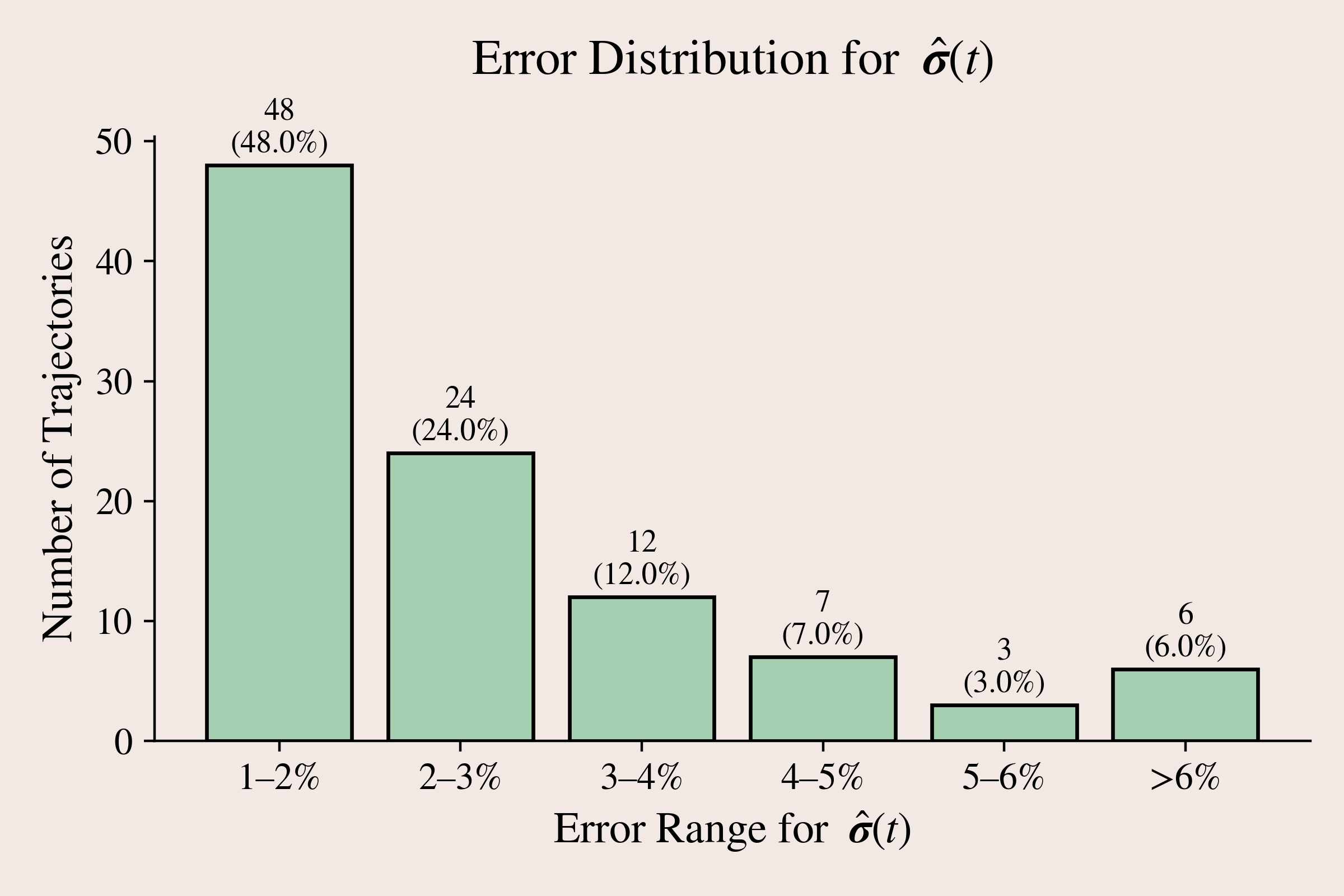}
	\caption{Analysis of homogenized stresses}
	\label{FIG:6}
\end{figure}
Figure~\ref{FIG:6} shows the analysis of the errors in homogenized stresses for trajectories in the test dataset. The analysis shows that about 94\% of the trajectories have an error less than 6\% in comparison to the reference values. Having demonstrated satisfactory accuracy in the predictions obtained with the hybrid model, we further evaluate its computational efficiency by comparing it against a Fortran solver. Our hybrid model computes the microscale quantities for a given trajectory in approximately 0.59 seconds, whereas the Fortran solver, executed using 16 MPI processes and 1 OpenMP thread per process, requires on average about 53 seconds for the same task.

\section{Summary and Future work}
In this contribution, we have developed a novel hybrid microscale model based on neural operators for multiscale simulations of rate-dependent materials. Our approach leverages a POD-DeepONet that takes macroscale strains and reduced internal variable representations as inputs to accurately predict the displacement field. By incorporating fundamental physical principles—including kinematic relations, constitutive laws, and evolution equations for internal variables—the model effectively reconstructs microscale strains, internal variables, and stresses. The hybrid model demonstrates high accuracy ($<6\%$) and significant computational efficiency ($\sim 100 \times$) compared to conventional methods. Importantly, this work presents the first hybrid model that integrates internal variables using a combination of surrogate and assistant strategies to predict microscale parameters within a multiscale simulation framework. This integration enables improved memory management by storing and reusing projection coefficients instead of full high-dimensional internal states at every macroscale integration point.

The promising results indicate that this hybrid approach can serve as a powerful tool for multiscale modeling, reducing computational costs while maintaining fidelity. Future work will focus on further exploiting the memory advantages of the reduced representation and extending the framework to arbitrary trajectories with variable time- step sizes.

\section*{Acknowledgments}
Funding from the German Federal Ministry of Education and Research (BMBF) for the project "MOdellkopplung im Kontext eines VIrtuellen Untertagelabors und dessen Entwicklungsprozess – MOVIE", within the research program "GEO:N – Geosciences for Sustainability" under the topic "Digital Geosystems: Virtual Methods and Digital Tools for Geoscientific Applications", is gratefully acknowledged [Reference: 03G0921A / 2024-01 to 2026-12].
Support by the Ministry of Science and Culture of Lower Saxony and the Volkswagen Foundation for the Research Training Group CircularLIB through the zukunft.niedersachsen program (MWK|ZN3678) is also acknowledged.

\vspace{1em}

\section*{Code Availability}

The training code used in this study is available at:  
\url{https://github.com/Dhananjeyan-Github/Hybrid-AI-Model-for-rate-dependent-simulations.git}

The dataset used to train the model has been archived and is publicly available on Zenodo:  
\url{https://doi.org/10.5281/zenodo.15676985}

This release includes only the training code and data. Evaluation and visualization routines are reserved for future work and will be released in a subsequent publication.

\vspace{1em}

\appendix
\renewcommand{\thefigure}{A.\arabic{figure}}
\setcounter{figure}{0}

\section*{Author Contributions}

\begin{itemize}
	\item \textbf{Dhananjeyan Jeyaraj} \href{https://orcid.org/0009-0000-5614-6197}{\texttt{0009-0000-5614-6197}}\\
	Conceptualization, Methodology, Software, Visualization, Formal Analysis, Investigation, Writing – Original Draft.
	
	\item \textbf{Hamidreza Eivazi} \href{https://orcid.org/0000-0003-3650-4107}{\texttt{0000-0003-3650-4107}}\\
	Conceptualization, Methodology, Software, Visualization, Writing – Review \& Editing.
	
	\item \textbf{Jendrik-Alexander Tröger} \href{https://orcid.org/0000-0002-4999-4558}{\texttt{0000-0002-4999-4558}}\\
	Conceptualization, Methodology, Writing – Review \& Editing.
	
	\item \textbf{Stefan Wittek} \href{https://orcid.org/0009-0007-3877-625X}{\texttt{0009-0007-3877-625X}}\\
	Supervision, Writing – Review \& Editing.
	
	\item \textbf{Stefan Hartmann} \href{https://orcid.org/0000-0003-1849-0784}{\texttt{0000-0003-1849-0784}}\\
	Supervision, Writing – Review \& Editing.
	
	\item \textbf{Andreas Rausch} \href{https://orcid.org/0000-0002-6850-6409}{\texttt{0000-0002-6850-6409}}\\
	Supervision, Writing – Review \& Editing.
\end{itemize}

\bibliographystyle{plainnat}
\bibliography{cas-refs}

\section{Appendix}

\begin{figure}[ht!]
	\centering
	\includegraphics[width=\textwidth]{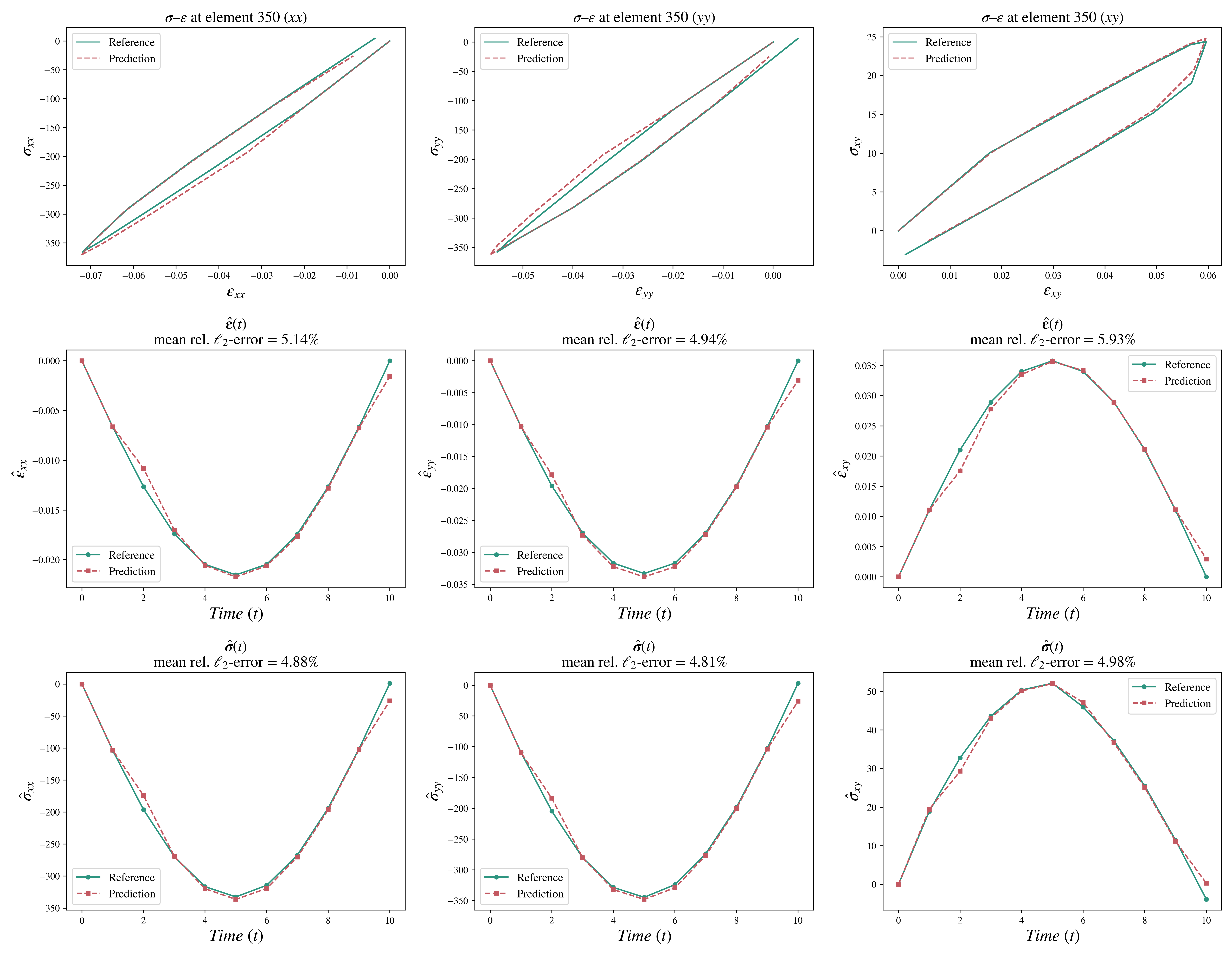}
	\caption{The first row presents the microscale stress–strain response for a single integration point exhibiting viscoelastic behavior. The second and third rows display the evolution of the homogenized strains and stresses, respectively, as functions of time. Stresses are reported in MPa and time is given in seconds.}
	\label{FIG:A1}
\end{figure}

The stress–strain response at a microscale integration point exhibiting viscoelastic behavior, along with homogenized strains and stresses over time as discussed in Subsection~\ref{sec:results}, are illustrated in Figure~\ref{FIG:A1}.

\end{document}